# Sudden modulation theory of hole spin-3/2 relaxation.

Yuri A. Serebrennikov

Qubit Technology Center

2152 Merokee Dr., Merrick, NY 11566

ys.qubit@gmail.com

We investigate the hole spin-3/2 relaxation process induced by nonadiabatic stochastic modulations of the instantaneous Luttinger Hamiltonian. The theory allows to consider fluctuations of both the direction and the magnitude of a hole wave vector in all regimes of momentum scattering: from collision-dominated to ballistic.

72.25.Rb  71.55.Eq

### I.  Introduction

Understanding of the hole spin relaxation process is important for fundamental physics and technological applications. Depending on the environment, the coherent lifetime of the hole spin-3/2 varies within the enormous range: from ultra-short < 1 ps in the bulk[1] to hundreds of microseconds in semiconductor quantum dots[2]. While hole-mediated ferromagnetism[3] and dynamic demagnetization of III-Mn-V materials[4] require ultra-fast hole spin relaxation, long spin decoherence is the prerequisite for realization of hole-based spintronic devices.  Since the valence band has *p*-symmetry the hole spin does not couple through Fermi contact interaction to lattice nuclei and, hence, is free from the hyperfine channel of decoherence, which is very efficient for electron spin at zero and low magnetic fields[5]. On the other hand, the intrinsic spin-orbit coupling (SOC) is much stronger for holes than for electrons. This leads to a strong mixing between the spin and orbital degrees of freedom of a carrier and provides the direct coupling of the resultant total angular momentum $\vec{J}$ with the hole lattice momentum $\vec{k}$. As a result, even in spherical bands and zero magnetic fields, $\vec{J}$ experiences a fast precession in the effective quadrupole field, which represents the anisotropic part of the instantaneous Luttinger Hamiltonian[6,7]. In bulk crystals, this field determines the splitting $\Delta_{HL}$ between heavy hole (HH) and light hole (LH) subbands. Both the magnitude and the direction of this field are modulated by thermal fluctuations of $\vec{k}$. Consequently, random scattering of a wave vector may induce nonadiabatic transitions between the HH and LH, which in turn lead to hole spin relaxation[6,7,8]. This process is qualitatively similar to Dyakonov-Perel (DP) mechanism of electron spin relaxation. Note, however, that an effective quadrupole field is orders of magnitude stronger than Rashba and Dresselhaus effective magnetic fields due to broken inversion symmetry[1]. This leads to an ultra-fast spin relaxation process in bulk crystals, usually on the scale of momentum relaxation time.

Most studies of the DP mechanism of spin relaxation were conducted within the collision-dominated regime, where the frequency of spin precession during the free path of the carrier is much less than the frequency of collisions[1]. In this limit, the spin motion turns into the process of small random walks in the angular space, which can be described within the Born-Redfield approximation (angular displacement of a spin is the small parameter of the stochastic perturbation theory). As already noticed, the strong SOC in the valence band leads to the fast precession of $\vec{J}$ in an effective quadrupole field, therefore, the usual approach based on the Born-Redfield approximation is not always legitimate for holes. In fact, the distinct optical orientation and relaxation of HH's and LH's that was clearly observed by Hilton and Tang[9] in undoped bulk GaAs requires a general theory that is able to describe the spin relaxation induced by stochastic modulation of the magnitude and direction of the wave vector outside the collision-dominated regime. The theory can be extended to the range of strong interactions only under certain assumptions regarding the random process. Very recently, we have formulated the novel approach to the problem based on the stochastic Liouville equation[10] that allows to calculate the response of $\vec{J}$ to sudden modulation of the direction of an effective quadrupole field[11]. The analytical solution of the problem describes the dumping of coherent oscillations from the collision-dominated ("fast motional") to ballistic ("slow motional") regime. The obtained results clearly demonstrate that the drastic drop in the rate of the hole spin relaxation process in low-dimensional semiconductor nanostructures in comparison to bulk crystals is related to suppression of the DP-like quadrupole mechanism by increased subband splitting and 2D-confinement of the hole motion.

The model investigated in Ref.[11] considers the stochastic wandering of the wave vector in the angular space as the sole source of *J*-relaxation. This approach closely resembles the description of the *nuclear* spin-3/2 relaxation. It disregards any effects related to stochastic modulation of the magnitude of $\vec{k}$ and ignores the inhomogeneous broadening due to the equilibrium distribution of $k^2$. Evidently, this model is not appropriate at finite temperatures. In this article, we extend the formalism of the sudden modulation theory of hole spin-3/2 relaxation. The extended theory allows to consider fluctuations of both the direction and the magnitude of an effective quadrupole field. We will show that the stochastic modulation of *k*-modulus opens an additional dephasing channel, which may lead to *J*-dephasing even in the absence of nonadiabatic stochastic jumps of the hole crystal momentum in the angular space. As expected, in the collision-dominated regime of a hole motion, the outcome of the calculations within the "strong collision" model presented here coincides with the result of the non-model stochastic perturbation theory[6]. In this regime, frequent scatterings drive carriers to the homogeneous state and consequently restrain both channels of *J*-relaxation.

## II. Theory

It has been shown in Ref.[6] that within the "spherical approximation"[12] the 4x4 matrix of the instantaneous Luttinger Hamiltonian[13] which represents $J = 3/2$ holes can be expressed in the following form (columns below correspond to m = 3/2, ½, -1/2, -3/2)

$$<3/2m_1; \vec{k}^{(M)} | H^{(M)}_{k^2} | 3/2m; \vec{k}^{(M)}> = \frac{\gamma_1 k^2}{2m_0}\delta_{mm_1} + \frac{\gamma_2}{m_0}\begin{pmatrix} D_k & 0 & \sqrt{3}E_k & 0 \\ 0 & -D_k & 0 & \sqrt{3}E_k \\ \sqrt{3}E_k & 0 & -D_k & 0 \\ 0 & \sqrt{3}E_k & 0 & D_k \end{pmatrix}. \quad (1)$$

Here the superscript (*M*) denotes the *principal-axes* system of the effective quadrupole tensor[8] $\ddot{Q}_{ij} = [L_i L_j + L_j L_i - (2L^2/3)\delta_{ij}]/2$, where $i,j = x_L, y_L, z_L$ represent the Cartesian basis in the space-fixed lab (*L*) frame, and $\vec{L}$ is the effective orbital angular momentum of a hole ($L = 1$), $\gamma_1$ and $\gamma_2$ are the dimensionless Luttinger parameters, $m_0$ is the bare electron mass, $D_k := -(2k_{z_M}^2 - k_{x_M}^2 - k_{y_M}^2)/2$, and $E_k := -(k_{x_M}^2 - k_{y_M}^2)/2$. This matrix can be represented in terms of the irreducible spin-tensor operators of the full rotation group[11] $T_{2q}(J) = (5/4)^{1/2}\sum_{m m_1} C^{Jm_1}_{Jm 2q} |Jm_1><Jm|$, $J = 3/2$ as

$$H^{(M)}_{k^2}(J) = H_0 + V^{(M)}_Q = (\gamma_1/2m_0)k^2 + (6)^{1/2}(\gamma_2/m_0)\sum_q (-1)^q K^{(M)}_{2q} T^{(M)}_{2-q}(J), \quad (2)$$

where $C^{2q}_{1\mu 1\mu_1}$ denotes the Clebsch-Gordon coefficient[14], $K^{(M)}_{20} = (2/3)^{1/2} D_k$, $K^{(M)}_{2\pm 1} = 0$, and $K^{(M)}_{2\pm 2} = E_k$. Here $H_0$ denotes the *isotropic* part of the kinetic energy of a hole, $V^{(M)}_Q$ stands for the effective quadrupole interaction, which represents the coupling between $\vec{J}$ and the lattice momentum in the *M*-frame and is clearly *anisotropic*, $[J^2, V^{(M)}_Q] \neq 0$. Physically this means that that the ballistic motion of a hole breaks the isotropy of the system and, similar to a crystal field, lifts the degeneracy of the $\Gamma_8$ ($|3/2\,m; k=0>$) "fine-structure" states that exists only at the $\Gamma$-point ($k=0$). In the axially symmetric case ($E_k = 0$) the matrix of $H^{(M)}_{k^2}$ is diagonal in the $|3/2\,m;\vec{k}>$ basis, $J_{z_M}$ is conserved, and the eigenfunctions of $H^{(M)}_{k^2}$ can be classified by the helicity $m = \hat{\vec{k}} \cdot \vec{J}^{(M)}$. Bands with $m = \pm 3/2$ correspond to HHs, while bands with $m = \pm 1/2$ represent LHs with the gap $\Delta_{HL} = 2\gamma_2 D_k / m_0$ determined by the magnitude of an effective quadrupole field. Due to the *T*-invariance of the problem (no magnetic interactions) each of these bands has Kramers degeneracy. Note that for holes moving along $z_M$, Eq.(2) can be written in the familiar form $H^{(M)}_{k^2} = (k^2/2m_0)[\gamma_1 + 2\gamma_2(J^2/3 - J^2_{Z_M})]$. Thus, even if the carrier equilibrium distribution in the *k*-space is isotropic, the instantaneous Luttinger Hamiltonian outside the zone center lacks spherical symmetry. Thermal motion of a charge carrier in the crystal results in the random modulation of $V_Q$, which connects the tightly coupled *L-S* subsystem (a hole) to the bath and is, therefore, responsible for interband transitions and $\vec{J}$-relaxation. The main advantage of the expansion (2) is the simplicity of the transformation of irreducible tensor operators $T_{2q}(J)$ under rotations of the coordinate system[14],

$$H^{(L)}_{k^2}(\Omega_t) = D^{-1}(\Omega_t) H^{(M)}_{k^2} D(\Omega_t) = \frac{\gamma_1}{2m_0}k^2 + \frac{\sqrt{6}\gamma_2}{m_0}\sum_{qp}(-1)^p T^{(L)}_{2p}(J) D^2_{q,-p}(\Omega_t) K^{(M)}_{2q}, \quad (3)$$

which significantly simplifies the theoretical study of the $\vec{J}$-relaxation process. Here $D(\Omega_t)$ is the operator of finite rotation, $\Omega_t = \{\alpha_t, \beta_t, \gamma_t\}$ is the set of Euler angles that represents the instantaneous orientation of the *L*-frame relative to the *M*-frame of reference at the moment *t*, $D^2_{q,-p}(\Omega_t)$ is the corresponding Wigner rotation matrix. The basic problem is the calculation of the response of $\vec{J}$ to a random realization of

$$V^{(L)}_Q(V,\Omega_t) = \frac{\sqrt{6}\gamma_2}{m_0}\sum_{qp}(-1)^p T^{(L)}_{2p}(J) D^2_{q,-p}(\Omega_t) K^{(M)}_{2q}, \quad (4)$$

where $V := \{D_k, E_k\}$. Without a precise definition of this process, this goal can be achieved only in the collision-dominated regime of hole motion (Born-Redfield approximation), where the stochastic perturbation is smaller than the inverse of the relevant correlation time. Henceforth, we shall assume that the main source of the stochastic time dependence of $V^{(L)}_Q$ is the stationary Markovian process in which the amplitude *V* of an effective quadrupole interaction and the orientation $\Omega$ of the M-frame varies suddenly at consecutive time moments $t_1, t_2, ...$ and is constant within the time-periods between them. The distribution of $t_i$ is of the Poisson type with the average time between "collisions" $\tau_Q$.

It is evident that such a process is a model. The time interval in which the variation of the Hamiltonian takes place ("collision" time) $\tau_c$ must be finite (~10 fs) even if short compared to $\tau_Q$. We can neglect the shortest of the times if this change is nonadiabatic $\|V_Q\|\tau_c \ll 1$ and, hence, the intricate details of a collision are unimportant. In this case, the orientation of $\vec{J}$ is the same immediately after the jump of $\vec{k}$ in the angular space. As a result, even if $\vec{J}$ is parallel to $\vec{k}$ during some interval $t_{i+1} - t_i$, (i.e., $\vec{J}$ commutes with $V_Q^{(L)}$), it will not commute with the Hamiltonian of the problem after a sudden change of $\Omega$ and begin to precess about the new direction of an effective quadrupole field. In general, $V_Q^{(L)}$ Eq.(4) depends on two multidimensional stochastic variables $\Omega$ and $V$. Since the former represents the instantaneous orientation of the $M$-frame in the angular space and the latter the strength of the coupling between $\vec{J}$ and the lattice momentum, they assumed to vary independently. This means that the equilibrium distribution of these variables is the product $\varphi(\Omega)\Phi(V)$, with each factor obeying the conservation of equilibrium $\varphi(\Omega) = \int f(\Omega,\Omega')\varphi(\Omega')d\Omega'$, $\Phi(V) = \int F(V,V')\Phi(V')dV'$. Here the quantities with and without the prime are values of the random variables before and after their change, respectively. The degree of correlation at the energy $\varepsilon_k$ is determined by the functions $f(\Omega,\Omega')$ and $F(V,V')$, which satisfy the normalization condition $\int f(\Omega,\Omega')d\Omega' = \int F(V,V')dV' = 1$. It then follows that in the Heisenberg representation an appropriately ensemble-averaged operator $J_{Z_L}^{(L)}(V,\Omega,t)$ obeys the stochastic Liouville equation of motion[10] ($\hbar = 1$):

$$\dot{J}_{Z_L}^{(L)}(V,\Omega,t) = i[H_{k^2}^{(L)}(V,\Omega), J_{Z_L}^{(L)}(V,\Omega,t)] \\ - \tau_Q^{-1}[J_{Z_L}^{(L)}(V,\Omega,t) - \int dV' F(V,V')\int d\Omega' f(\Omega,\Omega') J_{Z_L}^{(L)}(V',\Omega',t)] \quad (5)$$

This equation must be solved with the initial condition

$$J_{Z_L}^{(L)}(V,\Omega,t=0) = \Phi(V)\varphi(\Omega) J_{Z_L}^{(L)} \quad (6)$$

and the final physical information can be extracted from the ordinary integrals of the solution over $V$ and $\Omega$. Consequently, the autocorrelation function of $J_{Z_L}$ is given by

$$K_{J_{Z_L}}(t) = Tr \iint \rho_{eq}^{(L)}(V,\Omega) J_{Z_L}^{(L)}(V,\Omega,t) J_{Z_L}^{(L)} dV\, d\Omega, \qquad (7)$$

where $\rho_{eq}$ is the equilibrium density operator.

It is important to note here that due to the assumed isotropy of either bulk crystals or 2D nanostructures (in-plane isotropy), all directions of an effective quadrupole tensor are equiprobable. Therefore, the conditional probability density $f(\Omega,\Omega')$ depends only on the angle $\tilde{\Omega} = \Omega - \Omega'$ between the successive directions of the $M$-frame, i.e.,

$$f(\Omega,\Omega') = f(\tilde{\Omega}), \qquad (8)$$

and the density of its equilibrium angular distribution $\varphi(\Omega) = 1/8\pi^2$. If $f(\tilde{\Omega})$ and/or $F(V,V')$ are close to $\delta$-functions, the corresponding random values change negligibly at every jump, the process is strongly correlated ("weak collision" limit). If, on the other hand, an equilibrium distribution is re-established after every stochastic jump of the respective variable, i.e., $f(\tilde{\Omega}) = \varphi(\Omega)$ and $F(V,V') = \Phi(V)$, the corresponding process is uncorrelated ("strong collision" limit). The stochastic Liouville equation, Eq.(5), is rather general and provides the computational bridge between the spin relaxation of a charge carrier and fluctuations of the magnitude and direction of its crystal momentum (effective quadrupole field). This is a standard formulation of the problem in sudden modulation theory. It is mathematically closed, however, rather complex. Remarkably, the property of the kernel (8) is sufficient to advance in solving Eq.(5). As has been shown in Ref.[15], with respect to angular variables Eq.(5) can be reduced to a differential one, which is formally identical to the master equation of the impact theory (see Appendix):

$$\dot{\tilde{J}}_q^{(M)}(V,t) = i\hat{L}_{k^2}^{(M)}(V)\tilde{J}_q^{(M)}(V,t) - \tau_Q^{-1}[\tilde{J}_q^{(M)}(V,t) - \sum_{q_1}\hat{T}_{qq_1}\int F(V,V')\tilde{J}_{q_1}^{(M)}(V',t)\,dV']. \quad (9)$$

The initial condition to Eq.(9) is

$$\tilde{J}_q^{(M)}(V,t) = \Phi(V) J_q^{(L)^+}/3. \quad (10)$$

Here we introduce the following designations

$$\hat{L}_{k^2}^{(M)}(V)\tilde{J}_q^{(M)}(V,t) = [V_Q^{(M)}(V), \tilde{J}_q^{(M)}(V,t)], \quad (11)$$

$$\tilde{J}_q^{(M)}(V,t) = \int D(\Omega) J_q^{(L)}(V,\Omega,t) D(-\Omega) D_{q\,0}^1(\Omega)\,d\Omega, \quad (12)$$

$$J_{q=0} = J_{Z_L},\ J_{q=\pm1} = \mp(J_{X_L} \pm iJ_{Y_L})/\sqrt{2}. \quad (13)$$

Thus, the effect of random reorientation of $\vec{k}$ is reduced to a linear transformation of the $J_q$-components realized by the operator $\hat{T}_{qq_1}$ acting in the Liouville-space

$$\hat{T}_{qq_1}\tilde{J}_{q_1}^{(M)}(V,t) = \int f(\tilde{\Omega}) D(\tilde{\Omega}) \tilde{J}_{q_1}^{(M)}(V,t) D(-\tilde{\Omega}) D_{q\,q_1}^1(\tilde{\Omega})\,d\tilde{\Omega}. \quad (14)$$

The autocorrelation function Eq.(7) can be expressed via the solution of Eq.(9)

$$K_{J_{Z_L}}(t) = Tr\sum_q \int \rho_{eq}^{(M)}(V)\tilde{J}_q^{(M)}(V,t) J_q^{(L)}\,dV. \quad (15)$$

Equation (9) remains integral with respect to the two-dimensional amplitude of the perturbation. To make this equation solvable in principle, one must define the function $F(V,V')$. Let us assume that random changes of the magnitude of the effective quadrupole perturbation are non-correlated. In this case, the solution of Eq.(9) and the autocorrelation function Eq.(15) can be found by the routine numerical methods. Indeed, in the case of uncorrelated jumps of $V$ ("strong collisions") Eq.(9) is reduced to the following form

$$\dot{\tilde{J}}_q^{(M)}(V,t) = i\hat{L}_{k^2}^{(M)}(V)\tilde{J}_q^{(M)}(V,t) - \tau_Q^{-1}[\tilde{J}_q^{(M)}(V,t) - \Phi(V)\sum_{q_1}\hat{T}_{qq_1} <\tilde{J}_{q_1}^{(M)}(t)>], \quad (16)$$

where $<\tilde{J}_q^{(M)}(t)> = \int \tilde{J}_q^{(M)}(V',t)dV'$. Resolving Eq.(16) with respect to the Laplace-transformed operator $\tilde{J}_q^{(M)}(V,\omega) = \int_0^\infty \tilde{J}_q^{(M)}(V,t)\exp(-i\omega t)dt$ and integrating over $V$ we obtain the following compact result

$$<\tilde{J}_q^{(M)}(\omega)> = \sum_{q_1}\hat{R}_{qq_1}^{-1}(\omega)J_{q_1}^{(L)^+}/3, \quad (17)$$

where

$$\hat{R}_{qq_1}(\omega) = \hat{M}^{-1}(\omega)\delta_{qq_1} - \tau_Q^{-1}\hat{T}_{qq_1}, \quad (18)$$

$$\hat{M}(\omega) = \int \frac{\Phi(V)dV}{-i[\omega + \hat{L}_{k^2}^{(M)}(V)] + \tau_Q^{-1}}. \quad (19)$$

In the high-temperature limit, $V \ll k_B T$, one may ignore the dependence of the equilibrium density operator on $V$ and advance further in analytical solution of the problem. In this approximation the normalized spectral function

$$K_{J_{Z_L}}(\omega)/K_{J_L}(0) = \pi^{-1}\text{Re}\int_0^\infty K_{J_{Z_L}}(t)\exp(-i\omega t)dt/K_{J_L}(0)$$

can be expressed as follows

$$K_{J_{Z_L}}(\omega)/K_{J_L}(0) = (15\pi)^{-1}\text{Re}\,Tr\sum_{qq_1}J_q^{(L)}\hat{R}_{qq_1}^{-1}(\omega)J_{q_1}^{(L)^+}. \quad (20)$$

The integral in Eq.(19) can be taken with any normalized distribution function $\Phi(V)$, which permits certain freedom in physical modeling. Evidently, if we ignore the inhomogeneous broadening due to the equilibrium distribution of the magnitude of an effective quadrupole field, the kinetics of *J*-relaxation is determined solely by stochastic wandering of the wave vector in the angular space. This model corresponds to

$\Phi(V) = \delta(V - V_0)$ and has already been considered for spin-3/2 holes in Ref.[11]. It is formally equivalent to the well-studied problem of nuclear spin-3/2 relaxation induced by stochastic reorientation of the nuclear quadrupole Hamiltonian at zero magnetic fields[16]. It has been shown in Ref.[11] that if all holes are on the isotropic Fermi surface in the slow motional regime, $D_k^2 \tau_Q^2 >> 1$, Eqs.(17) - (20) describe the well resolved triplet structure in the spectral function, which reflects fast coherent oscillations of $\vec{J}$ in the effective quadrupolar field, $K_{J_{Z_L}}(t)/K_{J_{Z_L}}(0) = [3 + 2\cos(\Delta_{HL} t)]/5$, dumped by the relaxation process induced by nonadiabatic stochastic jumps of $\vec{k}$ in the angular space. The integral rate of this process, defined as $\tau_J(\Gamma_8) = \pi K_{J_0}(\omega \to 0)/K_{J_0}(0)$, is given by

$$\tau_J^{-1}(\Gamma_8) = \frac{14}{15} \tau_2^{-1} [1 + \frac{2\tau_4}{5\tau_2}]^{-1}, \qquad (21)$$

which does not depend on the magnitude of the effective quadrupole field. Here

$1/\tau_l = (1/\tau_Q) \int_{-1}^{1} f(\cos \tilde{\beta})[1 - P_l(\cos \tilde{\beta})] d(\cos \tilde{\beta})$ is the inverse orientational relaxation time of the *l*-rank tensor and $P_l$ is the Legendre polynomial. Note that if the term with cubic symmetry is included into the Luttinger Hamiltonian, the magnitude of an effective quadrupole field will depend on the direction of $\vec{k}$. This will lead to inhomogeneous broadening and may cause incomplete *J*-dephasing even in the pure ballistic, $\tau_Q \to \infty$, regime[8].

For finite values of $\tau_Q$, the thermal fluctuations of the *magnitude* of an anisotropic part of the instantaneous Luttinger Hamiltonian Eq.(4) will stochastically modulate the gap between HH and LH components of the $\Gamma_8$ quadruplet. This process leads to *J*-dephasing even if (hypothetically) the orientation of $\vec{k}$ is not affected by collisions, i.e., $f(\widetilde{\Omega}) = \delta(\widetilde{\Omega})$ and, hence, $\tau_2 = \tau_4 \to \infty$. Taking into account Eqs.(14) and (18), in this case, $\hat{R}_{qq_1}(\omega) = [\hat{M}^{-1}(\omega) - \tau_Q^{-1}]\delta_{qq_1}$ and Eq.(20) can be simplified further

$$K_{J_{Z_L}}(\omega)/K_{J_L}(0) = (15\pi)^{-1} \operatorname{Re} Tr \sum_q J_q^{(L)}[\hat{M}^{-1}(\omega) - \tau_Q^{-1}]^{-1} J_q^{(L)+}. \quad (22)$$

Note that in the absence of stochastic reorientations of the crystal momentum $V_Q^{(L)}$ does commute with itself at any instance of time. Therefore, fluctuations of *k*-modulus cannot induce HH-LH transitions and may cause only a *pure J-dephasing process*[17]. It follows from Eqs.(19) and (22) that the rate of this process strongly depends on $\Phi(V)$. It becomes negligible when $\Phi(V)$ is close to $\delta$-function.

In the collision-dominated regime, $\|V_Q^2\| \tau_Q^2 \ll 1$, the obtained solution Eqs.(17) - (20) can be expanded in a convergent series. Taking into account Eq.(2) and performing the sum over *m*-indexes in the first non-zero order of expansion in $D_k^2 \tau_Q^2$, we obtain the simple Lorentzian form of the spectral function Eq.(20), which corresponds to the exponential decay of the hole spin polarization with the rate[18]

$$\tau_J^{-1}(\Gamma_8) = \frac{8}{5}(\gamma_2/m_0)^2 <D_k^2> \tau_Q = \frac{2}{5}<\Delta_{HL}^2> \tau_Q, \quad (23)$$

where <...> denotes the average over the equilibrium distribution of the absolute values of the lattice momentum. We would like to stress here that, similar to DP mechanism of electron spin relaxation, the rate of *J*-dephasing Eq.(23) is proportional to $\tau_Q$ and is less than $1/\tau_Q$. As expected, Eq.(23) coincides with the non-model result of stochastic perturbation theory (see Ref.[6]).

### III. Conclusion

It should be clear from the above that as long as $\Delta_{HL}^2 \tau_c^2 \ll 1$ the decay of a hole spin polarization in bulk crystals is mainly due to nonadiabatic intersubband HH - LH transitions induced by stochastic reorientations of the hole crystal momentum, which strongly depend on the ratio between the $\Delta_{HL}$ and $\tau_Q$. In the collision-dominated regime of a hole motion, $\Delta_{HL}^2 \tau_Q^2 \ll 1$, the outcome of the calculations within the "strong collision" model presented here is consistent with the results of the non-model perturbation theory[6], which we would like to briefly summarize here. In the fast motional limit, the anisotropic part of instantaneous Luttinger Hamiltonian is *self-averaged* by rapid isotropic reorientations of $\vec{k}$. As a result, in bulk crystals the spherical symmetry of the system is restored, $J$ is a good quantum number, and it is impossible to distinguish between the HH and the LH components of the $\Gamma_8$ quadruplet. Nevertheless, the absence of HH optical orientation does not necessarily mean that the rates of angular and linear momentum relaxation are the same. According to Eq.(23) $1/\tau_J \sim \gamma_2^2 <k^4>$ and one may anticipate faster $J$-relaxation at higher temperatures and materials with larger $\gamma_2$. If we assume that at the room temperature $2\sqrt{<D_k^2>} \sim <k^2> \approx 10^{-3}/a_0^2$, where $a_0$ is the lattice constant[19], then for GaAs $\sqrt{<\Delta_{HL}^2>} \approx 40\, mEv$. For $\tau_Q = 80$ fs this gives $\tau_J \approx 0.3$ ps, which is approximately three times longer than the experimental value[9]. The dispersion of k-modulus and, thus, the amplitude of an effective quadrupole field at finite temperatures leads to inhomogeneous broadening of the spectral function Eq.(20). However, in the collision-dominated regime, frequent scatterings drive carriers to the homogeneous state and consequently suppress the relaxation. Since Hilton and Tang were able to distinguish between HH and LH bands[9], the system under study was outside the fast motional regime and, hence, the discrepancy between theoretical and experimental values of $\tau_J$ is not surprising. Apparently, when collisions are less frequent, $\Delta_{HL}^2 \tau_Q^2 \geq 1$, stochastic modulation of k-modulus open an additional dephasing channel (see Eq.(22)). It is important to note here that whereas increased subband splitting and confinement of the hole motion to 2D in low-dimensional nanostructures will suppress the former mechanism of spin relaxation[11], the latter may continue to operate. Recently, the hole spin dephasing due to stochastic modulation of the effective magnetic field (Rashba SOC term) in quantum wells with well separated LH and HH bands was studied by constructing and numerically solving the kinetic spin Bloch equations[20]. The study goes beyond the usual Born-Redfield approximation and takes into account the effects of the inhomogeneous broadening and Coulomb scattering. The detailed comparison with the results of this study for $\Delta_{HL}^2 \tau_Q^2 \geq 1$ and low hole concentration ($< 10^{11} cm^{-2}$) will be presented elsewhere.

**Appendix**.

Let us rewrite Eq.(7) in the following form

$$K_{J_{Z_L}}(t) := Tr \iint D(\Omega) \rho_{eq}^{(L)}(V,\Omega) J_0^{(L)}(V,\Omega,t) J_0^{(L)} D(-\Omega) d\Omega dV \qquad (A1)$$

Utilizing the transformation low

$$J_0^{(M)} = D(\Omega)J_0^{(L)}D(-\Omega) = \sum_q D_{q0}^1(\Omega) J_q^{(L)}, \tag{A2}$$

we can reduce Eq.(A1) to

$$K_{J_{Z_L}}(t) := Tr \sum_q \int \rho_{eq}^{(M)}(V) \tilde{J}_q^{(M)}(V,t) J_q^{(L)} dV, \tag{A3}$$

where we introduce the designation (see Eq.(12))

$$\tilde{J}_q^{(M)}(V,t) = \int D(\Omega) J_0^{(L)}(V,\Omega,t) D(-\Omega) D_{q0}^1(\Omega) d\Omega \tag{A4}$$

Note that operator $\tilde{J}_q^{(M)}(V,t)$ is determined in the $M$-frame. Now we shall see that the property of the kernel Eq.(8) is sufficient to derive the kinetic equation that is closed with respect to this operator. For that purpose, multiply the LHS of Eq.(5) by $D(\Omega)$ and its RHS by $D(-\Omega)D_{q0}^1(\Omega)$. Then integration over $\Omega$ yields

$$\dot{\tilde{J}}_q^{(M)}(V,t) = i\hat{L}_{k^2}^{(M)}(V)\tilde{J}_q^{(M)}(V,t) - \tau_Q^{-1}[\tilde{J}_q^{(M)}(V,t) - \\ - \int dV' F(V,V') \int d\Omega \int d\Omega' f(\Omega - \Omega') D(\Omega) J_0^{(L)}(V',\Omega',t) D(-\Omega) D_{q0}^1(\Omega)] \tag{A5}$$

Representing the integral term of Eq.(A5) as

$$\sum_{q_1} \int dV' F(V,V') \int d\Omega \int d\Omega' f(\Omega - \Omega') D(\Omega - \Omega')[D(\Omega') J_0^{(L)}(V',\Omega',t) D(-\Omega') D_{q_1 0}^1(\Omega')] \times \\ \times D(\Omega' - \Omega) D_{qq_1}^1(\Omega - \Omega')$$

and taking into account that integration over $\Omega$ is equivalent to integration over $\tilde{\Omega} = \Omega - \Omega'$ (Jacobian of this transformation equals 1), we obtain the following kinetic equation closed relative to the operator $\tilde{J}_q^{(M)}(V,t)$:

$$\dot{\tilde{J}}_q^{(M)}(V,t) = i\hat{L}_{k^2}^{(M)}(V)\tilde{J}_q^{(M)}(V,t) - \tau_Q^{-1}[\tilde{J}_q^{(M)}(V,t) - \\ - \int dV' F(V,V') \sum_{q_1} \int d\tilde{\Omega} f(\tilde{\Omega}) D(\tilde{\Omega}) \tilde{J}_{q_1}^{(M)}(V',t) D(-\tilde{\Omega}) D_{qq_1}^1(\tilde{\Omega})] \tag{A6}$$

In isotropic media $\varphi(\Omega) = 1/8\pi^2$, hence, the initial condition to this equation can be easily derived from Eq.(6), (A2), and (A4): $\widetilde{J}_q^{(M)}(V,t) = \Phi(V) J_q^{(L)^+}/3$.